\documentclass[aps,pre,twocolumn,superscriptaddress,showpacs]{revtex4-1}
\usepackage{amsmath,amssymb,graphicx,color}
\usepackage{bm}
\usepackage{bbm}
\def\Tr{\mbox{Tr}}

\begin{document}
\title{Generalized energy measurements and modified transient quantum fluctuation theorems}

\author{Gentaro Watanabe}
\affiliation{Asia Pacific Center for Theoretical Physics (APCTP), San 31, Hyoja-dong, Nam-gu, Pohang, Gyeongbuk 790-784, Korea}
\affiliation{Department of Physics, POSTECH, San 31, Hyoja-dong, Nam-gu, Pohang,
Gyeongbuk 790-784, Korea}
\author{B. Prasanna Venkatesh}
\affiliation{Asia Pacific Center for Theoretical Physics (APCTP), San 31, Hyoja-dong, Nam-gu, Pohang, Gyeongbuk 790-784, Korea}
\author{Peter Talkner}
\affiliation{Institut f\"{u}r Physik, Universit\"{a}t Augsburg, Universit\"{a}tsstra\ss e 1, D-86135 Augsburg, Germany}
\affiliation{Asia Pacific Center for Theoretical Physics (APCTP), San 31, Hyoja-dong, Nam-gu, Pohang, Gyeongbuk 790-784, Korea}

\date{\today}
\begin{abstract}
Determining the work which is supplied to a system by an external agent, provides a crucial step in any experimental realization of transient fluctuation relations. This, however, poses a problem for quantum systems, where the standard procedure requires the projective measurement of energy at the beginning and the end of the protocol. Unfortunately, projective measurements, which are preferable from the point of view of theory, seem to be difficult to implement experimentally. We demonstrate that, when using a particular type of generalized energy measurements, the resulting work statistics is simply related to that of projective measurements. This relation between the two work statistics entails the existence of modified transient fluctuation relations. The modifications are exclusively determined by the errors incurred in the generalized energy measurements. They are universal in the sense that they do not depend on the force protocol. Particularly simple expressions for the modified Crooks relation and Jarzynski equality are found for Gaussian energy measurements. These can be obtained by a sequence of sufficiently many generalized measurements which need not be Gaussian. In accordance with the central limit theorem, this leads to an effective error reduction in the individual measurements, and even yields a projective measurement in the limit of infinite repetitions.
\end{abstract}

\maketitle

\section{Introduction}\label{In}
In spite of the great theoretical as well as practical interest in the transient quantum fluctuation relations, their direct experimental confirmation is still missing.
These relations were pioneered by the classical fluctuation relations of Bochkov and Kuzovlev \cite{BK}, and are named after Jarzynski \cite{J} and Crooks \cite{C}.

Transient fluctuation relations restrict the statistics of work applied to a closed system by externally controlled, classical forces. The considered systems initially stay in thermal equilibrium but may be driven into regions of non-equilibrium far beyond the linear response regime. Yet, these relations yield key properties of equilibrium systems. Moreover, apart from their relevance for the understanding of the thermodynamics as well as the nonequilibrium behavior of small systems, the statistics of work, which is supplied to a system in a particular process, is of practical importance 
for the design and function of future nano-scale devices like engines and pumps \cite{BBM,SN,J13,DWLHG}.

According to the Crooks relation \cite{C} given by
\begin{equation}
p_\Lambda(w) = e^{-\beta (\Delta F -w)} p_{\bar{\Lambda}}(-w)\:,
\label{CRp}
\end{equation}
the probability density function (pdf), $p_\Lambda(w)$, of finding the work $w$ supplied to the system by a force $\lambda(t)$ varying in agreement with a prescribed protocol $\Lambda= \{\lambda(t)|0<t<\tau\}$, is connected with the time-reversed process.
This time reversed process 
is subject to the time-reversed protocol $\bar{\Lambda} =\{\lambda(\tau-t)|0<t<\tau\}$ \cite{tr}: It
starts at equilibrium, at the same inverse temperature $\beta$ as the forward process and at those parameter values $\lambda(\tau)$ that were finally reached in the forward process.  The quantity $\Delta F$ denotes the difference in free energy between the initial states of the forward and the backward processes and hence corresponds to the change in free energy of an isothermal process connecting $\lambda(0)$ with $\lambda(\tau)$.

In terms of the characteristic function of work, $G_\Lambda(u)$, which is the Fourier transform of the work pdf, $G_\Lambda(u) \equiv \int dw\, e^{iuw} p_\Lambda(w)$, the Crooks relation can equivalently be written as
\begin{equation}
G_\Lambda(u) = e^{-\beta \Delta F} G_{\bar{\Lambda}}(-u +i \beta)\:.
\label{CRG}
\end{equation} 
As an immediate consequence of the Crooks relation the Jarzynski equality follows. It expresses the mean value of exponentiated work in terms of the free energy difference $\Delta F$, reading
\begin{equation}
\langle e^{-\beta w} \rangle = e^{-\beta \Delta F}\:.
\label{JE}
\end{equation}

The validity of the transient fluctuation relation has been demonstrated 
for a wide variety of situations, including open classical \cite{JO} as well as closed \cite{K,Tas,TLH} and open \cite{CO,TCH,CTH} quantum systems, which also may be probed by measurements during the force protocol \cite{CTH10,CTH11,WVTCH}. Recent reviews are provided by \cite{JR,S,EHM,CHT}.
 
The main issue in the experimental confirmation of the quantum fluctuation relations is the determination of work. In the classical context, this does not present a basic problem because the work can be determined in an incremental way by integrating the supplied power which can be inferred from the instantaneous states of the system. Needless to say that, for this purpose, the system has to be continuously monitored. In quantum mechanics, the monitoring will have a severe impact on the system dynamics and also on the statistics of work. An additional difficulty comes from the fact that work is not an observable \cite{TLH}. 
Within the standard approach \cite{CHT}, the work supplied to a closed system by the action of a time-dependent force $\lambda(t)$ is expressed as the difference of the system energies at the time $\tau$ after a prescribed work protocol  has finished and at the starting time $t=0$ of the protocol.

Until now, in the literature, one may find  a proposal of a direct confirmation \cite{HSDL} and several elaborate ideas how to establish the transient fluctuation relations in a more indirect way \cite{DCHFGV,MDP,CBKZ}. Recently, an experimental realization of the proposal made in Ref.~\cite{MDP} was reported \cite{BSMASOGDPS}. These indirect methods \cite{DCHFGV,MDP,CBKZ,BSMASOGDPS} circumvent the measurement of work and infer its statistics by means of a simulation of the characteristic function of work, which is imprinted in the reduced state of an ancilla. Here we do not follow this procedure but investigate whether generalized energy measurements can be employed for determining the work.  

At the first glance, this does not seem to be a promising approach because 
we proved in a previous work \cite{VWT} that generalized measurements violate the transient fluctuation relations if they are not adapted to the actual force protocol in a special way. Only projective measurements can be universally used for arbitrary protocols \cite{VWT}. For the majority of generalized energy measurements, the resulting work statistics does not allow to determine the ``ideal'' work statistics obtained by use of projective measurements for the same force protocol.
However, for the special class of energy measurements introduced in  Section~\ref{Mdem}, the work statistics is connected to the ideal one such that the latter can be recovered from the former as discussed in Section~\ref{Ws}. In Section~\ref{Mfr} we derive modified fluctuation relations for this particular class of generalized energy measurements. 
Most importantly, the appearing modifications are completely determined by the error probabilities of the measurement devices but do not depend on the force protocol.
The form of the modification is such that these relations can directly be used to infer the changes of the system free energy. Particularly simple and easy to handle modifications result for Gaussian error distributions.      

In Section~\ref{Rem} we demonstrate how one may obtain measurements with Gaussian distributed errors from multiply repeated measurements with more general non-Gaussian
error distributions by using arguments that underlie the central limit theorem \cite{F}. A further beneficial effect of repeated measurements is a strong reduction of the error. The paper closes with Section~\ref{D}.

\section{Energy measurements}\label{Mdem}
We first collect the most important general properties of generalized measurements and then introduce  a special class of energy measurements. For a more complete account of the theory of generalized measurements we refer to the book by Wiseman and Milburn \cite{WM}.

\subsection{General properties of generalized measurements}
The description of generalized measurements in terms of positive operator valued measures (POVM) is very flexible. It not only allows for the assignment of a certain probability to find a pointer within some given range, but also determines the state of the system immediately after a measurement has been performed. The answers to both questions are given in terms of measurement operators $M_x$, where $x \in \mathcal{X}$ is a pointer state, $\mathcal{X}$ the totality of these states, and $M_x$ are bounded operators on the Hilbert space of the considered systems with a normalization condition specified below.

Once a value $x$ of the pointer is measured in a system, which stays in a state described by the density matrix $\rho$, the non-normalized postmeasurement state is given by $M_x \rho M^\dagger_x$. Both the normalization and the probability $p_x(\rho)$ to find $x$ in the state described by $\rho$ are given by
\begin{equation}
p_x(\rho) = \Tr\, M^\dagger_x M_x \rho\:.
\label{pxr}
\end{equation}
Because of the requirement that this probability should be normalized for any density matrix $\rho$, i.e., $\int_{\mathcal{X}} dx\, p_x(\rho) =1$,  the measurement operators must provide a partition of unity of the form
\begin{equation}
\int_{\mathcal{X}}dx\, M^\dagger_x M_x = \mathbbm{1}\:.
\end{equation}
Apparently, the eigenprojection operators of an observable, say, of the system Hamiltonian, define a proper set of measurement operators where the pointer values indicate the eigenstates of the observable. These then give rise to a projective measurement.

\subsection{Minimally disturbing energy measurements}
Next we introduce a particular class of energy measurements. 
As a special example of this class,  
we first consider a situation in which the energy measurements at the beginning and the end of the force protocol are performed by Gaussian superpositions of projective measurements of energy eigenstates, which can be expressed as
\begin{equation}
M_E(t) = \sum_n \frac{1}{(2 \pi \mu^2(t))^{1/4}} \exp \left ( \frac{-(e_n(t)-E)^2}{4 \mu^2(t)} \right ) \Pi_n(t)\:,
\label{MGSR}
\end{equation}
where $\Pi_n(t)$ and $e_n(t)$ denote the eigenprojection operators and the eigenvalues of the Hamiltonians $H(\lambda(t))$, respectively, with $t=0$ indicating the beginning and $t=\tau$ the end of the force protocol. Hence the spectral representation of these Hamiltonians is given by $H(\lambda(t)) = \sum_n e_n(t) \Pi_n(t)$. Calculating the conditional probability $q^{\text{Gauss}}_t(E|e_n(t))$ to measure the energy $E$ in the state $\Pi_n(t)/\Tr\, \Pi_n(t)$, one obtains a Gaussian distribution with mean value $e_n(t)$ and variance $\mu^2(t)$, reading
\begin{equation}
\begin{split}
q^{\text{Gauss}}_t(E|e_n(t)) &\equiv \Tr\, M^\dagger_E(t) M_E(t) \Pi_n/\Tr\, \Pi_n\\
&= \frac{1}{\sqrt{ 2 \pi \mu^2(t)}} e^{-(E-e_n(t))^2/(2 \mu^2(t))}\:.
\end{split}
\label{qG}
\end{equation}
The measurement operator defined by Eq.~(\ref{MGSR}) can be written in a compact form as
\begin{equation}
M_E(t) = q_t^{1/2}(E|H(t))\:.
\label{MEH}
\end{equation}
Any choice of the error pdf $q_t(E|H(t))$ leads to self-adjoint energy measurement operators  $M_E(t) = M^\dagger_E(t)$ which therefore, according to the Wiseman-Milburn taxonomy, are {\it minimally disturbing} \cite{WM}.
In the example of a Gaussian energy measurement operator (\ref{MGSR}),
the error distribution is a function of the difference $E-e_n(t)$ only but does not depend on $E$ and $e_n(t)$ separately, and hence
\begin{equation}
M_E(t) = q^{1/2}_t(E \mathbbm{1} -H(t)|0)\:.
\label{Mq}
\end{equation}
We call minimally disturbing energy measurements of this type {\it homogeneous}. 

Finally, we conclude that the Gaussian measurement operator (\ref{MGSR}) is homogeneous because the deviation of the average from the condition, $\langle E \rangle_{q_t} - e_n(t)$, is independent of $n$ --- in fact, it vanishes --- and the variance $\mu^2(t)$ is independent of the condition $e_n(t)$. Here $\langle E \rangle_{q_t} $ and  $\mu^2(t)$  are the average and variance determined by the pdf $q_t(E|e_n(t))$, respectively.

\section{Work statistics with minimally disturbing energy measurements}\label{Ws}
As described in Introduction, the work statistics is based on the measurement of energies $E$ and $E'$ at the beginning and the end of the force protocol, respectively, and is hence determined by the joint probability $P_\Lambda(E',E)$, which, for general energy measurement operators $M_E(t)$ of the type of Eq.~(\ref{MEH}), is given by
\begin{equation}
\begin{split} 
P_\Lambda(E',E) &= \Tr\, M^2_{E'}(\tau) U(\Lambda) M_E(0) \rho_0M_E(0) U^\dagger(\Lambda) \\
&= \sum_{m,n} q_\tau(E'|e_m(\tau)) q_0(E|e_n(0)) p_\Lambda(m,n)\:,
\label{PEE}
\end{split}
\end{equation}  
where we allow for different measurement operators for the initial and final energy measurements, characterized by conditional pdfs $q_0(E|e_n(0))$ and $q_\tau(E'|e_m(\tau))$. 
The operator $U(\Lambda)\equiv U_{\tau,0}$ governs the time evolution from the beginning to the end of the force protocol and follows  as the solution of the Schr\"odinger equation,
\begin{equation}
i \hbar \partial U_{t,s} / \partial t = H(\lambda(t)) U_{t,s}\: ,
\label{SE}
\end{equation}
with the initial condition 
\begin{equation}
U_{s,s} = \mathbbm{1}\:.
\label{iq}
\end{equation} 
In going to the second line of Eq.~(\ref{PEE}), we interchanged the order of the initial measurement operator $M_E(0)$ and the initial density matrix $\rho_0 = Z^{-1}(0) e^{- \beta H(\lambda(0))}$, where $Z(0)=\Tr\, e^{-\beta H(\lambda(0))}$. This is possible because both operators are functions of the same Hamiltonian $H(\lambda(0))$. 
Here, $p_\Lambda(m,n)$ denotes the joint probability to find the eigenstates $n$ and $m$ in projective energy measurements at the beginning and the end of the force protocol, respectively. It reads
\begin{equation}
p_\Lambda(m,n) = \Tr\, \Pi_m(\tau) U(\Lambda) \Pi_n(0) \rho_0 U^\dagger(\Lambda)\:.
\label{pnm}
\end{equation}  
The work pdf $p_\Lambda(w)$ can be expressed in terms of the joint probability $P_\Lambda(E',E)$ as
\begin{equation}
p_\Lambda(w) = \int dE dE'\, \delta(w-E'+E) P_\Lambda(E',E)\:,
\label{pw}
\end{equation}
which leads to the following expression for the characteristic function $G_\Lambda(u)$:
\begin{equation}
\begin{split}
G_\Lambda(u)&= \int dw\, e^{i u w} p_\Lambda(w)\\
&= \sum_{m,n} \int dE dE'\, e^{i u(E'-E)}  \\
&\quad \times q_\tau(E'|e_m(\tau)) q_0(E|e_n(0)) p_\Lambda(m,n) \\
&= \sum_{m,n} g_\tau(u|e_m(\tau)) g_0(-u|e_n(0)) p_\Lambda(m,n)\:,
\label{Ggg}
\end{split}
\end{equation}      
where $g_t(u|e) \equiv \int dE\, e^{i u E}q_t(E|e)$ is  the characteristic functions of the conditional pdf $q_t(E|e)$ with $t=0,\:\tau$.

The characteristic function $G_\Lambda(u)$ takes a considerably simple form for homogeneous energy measurements. As the Fourier transform of a shifted function, $q_t(E|e) = q_t(E-e|0)$, the characteristic function of the measurement error becomes
\begin{equation}
g_t(u|e) = e^{iu e} g_t(u)\:,
\label{gtue}
\end{equation}
where 
\begin{equation}
g_t(u) \equiv \int dE\, e^{i u E} q_t(E|0)\:.
\label{gtu}
\end{equation}
Putting the expression (\ref{gtue}) into the work characteristic function, one finds that it is represented by a product of a protocol-independent function of $u$ and the characteristic function for projective measurements, hence, reading 
\begin{equation}
G_\Lambda(u) = g_\tau(u) g_0(-u) G^{\text{proj}}_\Lambda(u)
\label{GGp}
\end{equation}
with the characteristic function for projective energy measurements given by
\begin{equation}
 G^{\text{proj}}_\Lambda(u) = \sum_{m,n} e^{i u (e_m(\tau) -e_n(0))} p_\Lambda(m,n)\:.
\label{Gp}
\end{equation}
The work pdf can then be expressed as a convolution of the projective work pdf
with a protocol-independent pdf describing the combined effect of the errors incurred in the initial and final energy measurements. It takes the form
\begin{equation}
p_\Lambda(w) = \int dE\, Q(E) p^{\text{proj}}_\Lambda (w-E)\:.
\label{pQ}
\end{equation}
The combined measurement error pdf $Q(E)$ is given by
\begin{equation}
Q(E) = \int dy\, q_\tau(E+y|0) q_0(y|0)\:.
\label{Qqq} 
\end{equation}
On the other hand, one can show that the work pdf based on minimally disturbing energy measurements is given by the convolution of the projective work pdf with a protocol-independent error distribution only if the measurements are homogeneous in the sense of Eq.~(\ref{Mq}). If the measurement error distributions of the first and the second measurements are known, the work pdf resulting from a projective energy measurement can be reconstructed.

Using the characteristic function of a Gaussian distribution given by 
\begin{equation}
g(u)= \int dw\, e^{i w u} \frac{1}{\sqrt{2 \pi \mu^2}} e^{-w^2/(2\mu^2)} = e^{-\mu^2 u^2/2}\:,
\label{gG}  
\end{equation}
one finds the work characteristic function to become
\begin{equation}
G_\Lambda(u) = e^{-(\mu^2(\tau) + \mu^2(0)) u^2/2} G^{\text{proj}}_\Lambda(u)
\label{GG}
\end{equation}
for energy measurements with homogeneous Gaussian error distributions.
Accordingly, the work pdf  
results from the projective work pdf convoluted with a Gaussian. 
With Eq.~(\ref{pQ}) it becomes
\begin{equation}
\begin{split}
p_\Lambda(w) &= \int \frac{dE}{\sqrt{2 \pi (\mu^2(\tau) + \mu^2(0))}}\\ &\quad \times e^{-(w-E)^2/[2(\mu^2(\tau) + \mu^2(0))]}\, p^{\text{proj}}_\Lambda(E)\:.
\end{split}
\label{pGp}
\end{equation}
The effective error distribution, $Q(E) = \left [2 \pi (\mu^2(\tau) + \mu^2(0)) \right ]^{-1/2} e^{-E^2/[2(\mu^2(\tau) + \mu^2(0))]} $, describing the combined disturbance in the first and the second energy measurement, is also a Gaussian with vanishing mean value. Its variance is given by the sum of the variances of the initial and final Gaussian error distributions.

\section{Modified fluctuation relations}\label{Mfr}
Putting $u=i\beta$ in the expression (\ref{GGp}) for the characteristic function of work, we obtain on the left-hand side $G_{\Lambda}(i\beta)= \langle e^{-\beta w} \rangle$. The right-hand side can be expressed by means of the Jarzynski equality, which holds for projective energy measurements, leading to a modified Jarzynski equality for homogeneous, minimally disturbing energy measurements 
of the form \cite{note_interrupt}
\begin{equation}
\begin{split}
\langle e^{-\beta w} \rangle &= g_\tau(i \beta) g_0(- i \beta) e^{-\beta \Delta F}\\
&= \langle e^{-\beta E}\rangle_{q_\tau} \langle e^{\beta E}\rangle_{q_0}  e^{-\beta \Delta F}\:,
\end{split}
\label{MJE}
\end{equation}
where, in the second line, we expressed the characteristic functions of the error distributions at $u=\pm i \beta$ by the mean values of the exponentiated energy with respect to the corresponding error distributions using the notation $\langle \cdot \rangle_{q_t} = \int dE \cdot q_t(E|0)$.
We want to emphasize that the correction factor $\langle e^{-\beta E} \rangle_{q_\tau} \langle e^{\beta E} \rangle_{q_0}$ is protocol independent. For Gaussian
error distributions (\ref{qG}) it becomes
\begin{equation}
\langle e^{-\beta E} \rangle_\tau \langle e^{\beta E} \rangle_0 = e^{(\mu^2(\tau) + \mu^2(0))\beta^2/2}\:.
\label{cG}
\end{equation}
Note that the modification to the original Jarzynski equality
is particularly simple in this case. As long as we know the variance of
the Gaussian error distribution, we can determine the free energy change 
using the modified Jarzynski equality (\ref{MJE}) in a similar manner
to the original case, except for the extra numerical factor given
by Eq.~(\ref{cG}).

Next we discuss a modified Crooks relation for homogeneous, minimally disturbing energy measurements. Starting from the  Crooks relation (\ref{CRG}), holding for projective measurements, we may express the characteristic functions for projective measurements by means of Eq.~(\ref{GGp}) in terms of those for homogeneous, minimally disturbing measurements. In this way we obtain a modified Crooks relation of the form \cite{note_interrupt}
\begin{equation}
G_\Lambda(u) = \frac{g_\tau(u) g_0(-u)}{g_0(-u+i\beta ) g_\tau(u - i \beta)} e^{-\beta \Delta F} G_{\bar{\Lambda}}(-u+ i\beta)\:.
\label{MCRG} 
\end{equation}
By multiplying both sides by $g_0(-u+i\beta ) g_\tau(u - i \beta)$ and performing the inverse Fourier transform, one obtains the following modified Crooks relation in terms of the work pdfs reading
\begin{equation}
\begin{split}
&\int dE\, e^{\beta E} Q(E) p_\Lambda(w-E) = e^{-\beta (\Delta F -w)}\\
&\times \int dE\, e^{-\beta E} Q(E) p_{\bar{\Lambda}}(E-w)\:,
\end{split}
\label{MCRp}
\end{equation}
where the total error pdf $Q(E)$ is given by Eq.~(\ref{Qqq}). Either of the two equivalent variants (\ref{MCRG}) and (\ref{MCRp}) of the modified Crooks relation can be employed for determining the free energy change $\Delta F$ provided, the total error pdf is known.

For a Gaussian error distribution (\ref{qG}), the modified Crooks relation in terms of the characteristic function, Eq.~(\ref{MCRG}), reduces to 
\begin{equation}
G_\Lambda(u) = e^{-\beta (\mu^2(\tau)+\mu^2(0))(i u + \beta/2)} e^{-\beta \Delta F} G_{\bar{\Lambda}}(-u+i\beta)\:.
\label{MCGG}
\end{equation}   
In this case, the inverse Fourier transformation leads to a more direct relation between the forward and the backward work pdfs without involving integral transformations. 
The result can be brought in the following form:
\begin{equation}
\begin{split}
&p_\Lambda\left(w-\beta (\mu^2(\tau) + \mu^2(0))/2\right) = e^{-\beta(\Delta F -w)}\\ 
&\quad \times p_{\bar{\Lambda}}\left(-w-\beta (\mu^2(\tau) + \mu^2(0))/2\right)\:.
\end{split}
\label{MCRpG}
\end{equation}
The modification from the original form of the Crooks relation consists of a shift in the arguments of the forward and the backward work pdfs by the product of $\beta$ and the arithmetic mean of the variances of the error distribution of the energy measurements in the beginning and the end of the protocol. The exponential factor connecting the forward and backward work pdfs remains the same as in the original Crooks relation. 

In the case of Gaussian energy measurements, inferring free energy differences from experimentally determined work pdfs by means of the modified Crooks relation appears to be considerably simpler than for an arbitrary homogeneous, minimally disturbing measurement: For Gaussian measurements the arguments of the work pdfs only have to be shifted in a specific way, whereas for more general homogeneous measurements the forward and the backward work pdfs must be convoluted with functions depending on the effective error distributions.

\section{Repeated energy measurements}\label{Rem}

In the previous sections we found that minimally disturbing energy
measurements lead to modified fluctuation theorems. The resulting
modifications are solely determined by the characteristic functions
$g_0(u)$ and $g_\tau(u)$, specifying the errors of the initial and
final energy measurements. A particularly simple form of the
modification emerges for generalized measurements with Gaussian
distributed errors.  Using the central limit theorem we demonstrate in
the present section that a frequent repetition of initial and final
energy measurements leads to a modified fluctuation theorem of the
same form as it would result from a single pair of Gaussian energy
measurements. This result is universal for all repeated energy
measurements having homogeneous and minimally disturbing error
distributions with finite variance values.

Hence we consider a situation where the energy in the beginning and at the end is not only measured once but instead, several times. We suppose that the individual measurements are homogeneous and minimally disturbing energy measurements which are characterized by conditional error pdfs $q_t(E|e_n(t)) = q_t(E-e_n(t)|0)$, which may be different in the beginning ($t=0$) and at the end ($t=\tau$) of the protocol, but within these two sets of measurements they are supposed to be identical. We shall comment on generalizations later on.

We assume, as we already did implicitly, that the measurements are short on the time scale of the unitary dynamics of the system such that they can be considered instantaneous, though we may allow for some time elapsing between two subsequent measurements. We only require that all energy measurements performed at the beginning take place before the protocol has started and after the system has equilibrated and is isolated from the thermal bath. 
The second set of measurements at the end of the protocol must be performed after the force parameter has reached its final value $\lambda(\tau)$.
For such an experimental setup, the joint probability $P_\Lambda(\mathbf{E}',\mathbf{E} )$ to find the energies $\mathbf{E} = \{E_1, E_2, \ldots, E_N \}$ prior to the start of the protocol and  $\mathbf{E}' = \{E'_1, E'_2, \ldots, E'_N \}$   after the protocol has ended is given by 
\begin{equation}
\begin{split}
P_\Lambda(\mathbf{E}',\mathbf{E} ) &=\Tr\, \mathcal{M}'_{\mathbf{E}'} U(\Lambda) \mathcal{M}_{\mathbf{E}} \rho_0 \mathcal{M}^\dagger_{\mathbf{E}} U^\dagger(\Lambda) \mathcal{M}'^\dagger_{\mathbf{E}'}    \\
&= \Tr\, \mathcal{M}'^\dagger_{\mathbf{E}'} \mathcal{M}'_{\mathbf{E}'} U(\Lambda) \mathcal{M}_{\mathbf{E}}\rho_0  \mathcal{M}^\dagger_{\mathbf{E}} U^\dagger(\Lambda)\:.
\label{pNENE}
\end{split}
\end{equation}
In going to the second line we made use of the invariance of the trace under cyclic permutations. The operators 
\begin{equation}
\begin{split}
\mathcal{M}'_{\mathbf{E}'} &= \prod^{N-1}_{k=0} \left [M_{E'_{N-k}}(\tau) U(\lambda(\tau)) \right ]\:,\\
\mathcal{M}_\mathbf{E} &= \prod^{N-1}_{k=0} \left [M_{E_{N-k}}(0) U(\lambda(0)) \right ]
\end{split}
\label{MM}
\end{equation}
describe the collective action of measurements after and before the protocol, respectively.
The time-evolution operators $U(\lambda(t))= e^{-i \epsilon H(\lambda(t)) /\hbar}$, with $t=0,\tau$, propagate the state between two subsequent measurements separated by the time $\epsilon$. 
In principle, the products in Eq.~(\ref{MM}) are ordered  with decreasing indices labeling the sequence of measurements from the left to the right. However, because both the measurement operators and the time evolution operators are functions of the same Hamiltonian, all factors of each product mutually commute with each other.
Moreover, the initial density matrix $\rho_0$ as a function of $H(\lambda(0))$ commutes with the operator $\mathcal{M}_{\mathbf{E}}$. As a consequence, in both products $\mathcal{M}'^\dagger_{\mathbf{E}'} \mathcal{M}'_{\mathbf{E}'}$ and 
$\mathcal{M}_{\mathbf{E}} \mathcal{M}^\dagger_{\mathbf{E}}$ the time evolution operators combine to unit operators and the expressions yield
\begin{equation}
\begin{split}
P_\Lambda(\mathbf{E}',\mathbf{E} ) &= \Tr \prod_{k=1}^N M^2_{E'_k}(\tau) U(\Lambda) \prod_{k=1}^N M^2_{E_k}(0) \rho_0 U^\dagger(\Lambda)\\
& = \sum_{m,n} \prod_{k=1}^N q_\tau(E'_k|e_m(\tau)) q_0(E_k|e_n(0)) p_\Lambda(m,n)\:.
\end{split}
\label{Ppmn}
\end{equation} 
Taking as  estimates of the work the difference of the arithmetic means of the energies measured after the end and before the beginning of the protocol, i.e., 
estimating the work as $w= N^{-1} \sum_k (E'_k - E_k)$, 
we obtain for the work pdf, the expression
\begin{equation}
\begin{split}
p_\Lambda(w)& = \int d^N \mathbf{E}\: d^N \mathbf{E}'\: \delta \!\left (w-\frac{1}{N}\sum_{k=1}^N (E'_k -E_k)\right )\\
&\quad 
\times P_\Lambda(\mathbf{E}',\mathbf{E} )\:.
\end{split}
\label{pP}
\end{equation}
Using Eq.~(\ref{Ppmn}) the corresponding characteristic function can be further evaluated to yield
\begin{equation}
\begin{split}
G_\Lambda(u) &= \sum_{n,m} \prod_{k=1}^N g_\tau(u/N|e_m(\tau))  g_0(-u/N|e_n(0))\\
&\quad \times p_\Lambda(m,n)\:.
\end{split}
\label{Gggp}
\end{equation}
For homogeneous measurements this expression can be further simplified with the help of Eq.~(\ref{gtue}), yielding
\begin{equation}
G_\Lambda(u) = \left [g_\tau\left (\frac{u}{N} \right) g_0\left (-\frac{u}{N} \right) \right ]^N G^{\text{proj}}_\Lambda(u)\:,
\label{GggGp}
\end{equation}
where $G_{\Lambda}^{\text{proj}}$ is the characteristic function for projective energy measurements defined in Eq.~(\ref{Gp}).
The prefactor in front of the characteristic function for projective measurements coincides with the characteristic function of a sum of $N$ independent, identically distributed random numbers. 
Under the condition that the second moments of the error distributions $q_t(E|0)$ exist for $t=0,\tau$, the rationale of the central limit theorem applies \cite{F}. Then the cumulant generating functions $k_t(u) = \ln g_t(u)$ 
can be written as $k_t(u) = -\mu^2(t) u^2/2 + o(u^2)$, where $\mu^2(t) = \int dE\:E^2 q_t(E|0)$, and $\lim_{x \to 0} o(x)/x =0$. Consequently, for large numbers of energy measurements, Eq.~(\ref{GggGp}) tends to
\begin{equation}
G_\Lambda(u) = e^{-(\mu^2(\tau) + \mu^2(0)) u^2/(2N)} G^{\text{proj}}_\Lambda(u)\:,
\label{GGG}
\end{equation}
and therefore the total error pdf $Q_N(E)$ becomes  Gaussian with the variance $\mu^2_{\text{eff}} \equiv (\mu^2(\tau) + \mu^2(0))/N$. 
For finite numbers of measurements, the resulting error distribution will deviate from a Gaussian most pronouncedly at the tails of the distribution. 
The details of these deviations depend on the combined error pdf $\int dy\, q_\tau(E+y) q_0(y)$ of single pairs of initial and final energy measurements and can be estimated for large numbers of measurements by the corresponding rate function of large deviations \cite{E}.

Alternatively, the deviations of the combined error pdf from a Gaussian pdf can be quantified by its cumulants $\kappa_n$, which can be determined from the cumulant generating function 
\begin{equation}
k(u) = N\, \left[k_\tau(u/N)+ k_0(u/N) \right] = \sum_n \frac{\kappa_n}{n!} (iu)^n
\label{ku}
\end{equation}  
by an $n$-fold differentiation with respect to $iu$, i.e., as $\kappa_n = d^n k(u)/d(iu)^n|_{u=0}$ \cite{vK}. For the sake of simplicity, we assume that cumulants of all orders $n$ exist. The Gaussian is ruled by the first two cumulants $\kappa_1$, which, in the present case, vanishes because it coincides with the average error, and $\kappa_2= (\mu^2(\tau) + \mu^2(0))/N$, which agrees with the variance. The presence of higher cumulants indicates deviations from a Gaussian distribution.  The third-order cumulant indicates the skewness and the fourth-order the kurtosis, indicating whether the tails of the pdf contain more ($\kappa_4 >0$) or less weight ($\kappa_4 <0$), compared to a Gaussian.
Because of the particular scaling of the cumulant generating function (\ref{ku}) with the number $N$ of measurements, the cumulants themselves depend on $N$ as
\begin{equation}
\frac{\kappa_n}{(\kappa_2)^{n/2}} = c_n N^{1-\frac{n}{2}}\:, \qquad n\geq 2\:,
\label{kappa}
\end{equation}
where we used the square root of the variance $\kappa_2$ as the typical size of the error for comparison. The coefficients $c_n$ depend on the particular error distributions $q_t(E|0)$, $t = 0, \tau$ and may grow  with the order $n$ of the cumulants faster than $a^{n/2}$, where $a$ is a constant with $|a|>1$. To achieve relatively small cumulants of high order will then require a larger number $N$ of measurements than is needed to control the lower orders such as $\kappa_3$ and $\kappa_4$. 

A Gaussian distribution is also approached if the error distributions of the individual energy measurements are different from each other but still have a finite variance. 
In this case, the variance of the error pdf $Q_N(E)$ is given by
\begin{equation}
\mu^2_{\text{eff}} =\frac{1}{N^2} \sum_{k=1}^N (\mu^2_k(\tau) +\mu^2_k(0))\:,
\label{mueff}
\end{equation}
where $\mu^2_k(t)$ is the variance of the error pdf $q^{(k)}_t(E|0)$ of the $k$th measurement.    

Finally, we want to stress that infinitely many homogeneous and minimally disturbing energy measurements result in a projective measurement. For a large, but finite number, the estimated energy becomes Gaussian distributed with a variance that is proportional to the inverse of the number of measurements.

\section{Conclusions}\label{D}
From our earlier work \cite{VWT}, we know that in general replacing initial and final projective energy measurements by generalized measurements leads to work pdfs which are not compatible with the transient quantum fluctuation theorems. 
In the present work, we found that, within the  class of minimally disturbing energy measurements, i.e., for measurements that are described by self-adjoint functions of the Hamiltonian representing the energy,
so-called homogeneous measurements lead to rather simple modifications of the work statistics. 
We recall that a measurement is homogeneous if the probability of finding the energy $E$ in an energy eigenstate with the eigenvalue $e_n$ depends only on the difference $E-e_n$, i.e., the conditional error pdf $q(E|e_n)$ is invariant under a common shift of its arguments. For these measurements the work pdf can be expressed by a convolution of the corresponding pdf for projective energy measurements with an error distribution solely determined by the error distributions of the energy measurements. This leads to modified fluctuation relations.   
Beyond their mere existence, the remarkable property of these modified fluctuation relations lies in the fact that all aspects in which they deviate from the standard fluctuation relations are independent of the particular protocol and can be expressed in terms of the error distributions imposed by the measurements. If the error distributions of the initial and final energy measurements are known, these modified fluctuation relations can be used to determine the free energy change.

The modifications are particularly simple for Gaussian energy measurements. 
These can be obtained from repeatedly applied {\it arbitrary} homogeneous and minimally disturbing energy measurements, provided their error distributions have well-defined variance values. As a consequence of the central limit theorem, the resulting error probability of the arithmetic mean of the measurement results approaches a Gaussian distribution, which converges to a $\delta$ function in the limit of infinitely many measurements, hence yielding a projective measurement.    

\begin{acknowledgments} 
We acknowledge the Max Planck Society, and the Korea Ministry of Education, Science and Technology (MEST), Gyeongsangbuk-Do, Pohang City, for the support of the JRG at APCTP. We are also supported by the Basic Science Research Program through the National Research Foundation of Korea, funded by the MEST (No. 2012R1A1A2008028). 
\end{acknowledgments}


\begin{thebibliography}{99}
\bibitem{tr} For the sake of simplicity we restrict ourselves to forces which transform evenly under time reversal. The inclusion of odd forces is straightforward and does not change any of the results presented in this paper.
\bibitem{note_interrupt} The generalized fluctuation theorems
  of Eqs.~(\ref{MJE}) and (\ref{MCRG}) can hold even when the
  force protocol is interrupted by a generalized measurement of an
  arbitrary observable. For the modified Jarzynski equality (\ref{MJE}) to hold,
  the necessary and sufficient condition on the intervening
  measurement operator is given by Eq.~(15) of Ref.~\cite{WVTCH}.
  Regarding the modified Crooks relation (\ref{MCRG}), only sufficient
  conditions are known. The modified Crooks relation (\ref{MCRG})
  holds if the intervening measurement operator satisfies Eq.~(23) of
  Ref.~\cite{WVTCH} or Eqs.~(24) and (25) of Ref.~\cite{WVTCH}.
\end{thebibliography}
\end{document}